\documentclass[aps,twocolumn,floatfix,epsfig]{revtex4}
\usepackage{pdfpages}
\usepackage{epsf}
\usepackage{subfigure}
\usepackage{amsmath}
\usepackage{epstopdf}
\usepackage{mathrsfs}
\usepackage{natbib}
\usepackage{amssymb,latexsym}
\usepackage{dcolumn}
\usepackage{graphicx}

\def\be{\begin{equation}}
\def\ee{\end{equation}}
\def\ba{\begin{eqnarray}}
\def\ea{\end{eqnarray}}

\graphicspath{{images/}}
\begin{document}
\title{\large \bf  Gravitational collapse in the AdS background and the black hole formation }
\author{Alireza Allahyari}
\affiliation{Department of Physics, Sharif University of Technology,
Tehran, Iran }
\email{allahyari@physics.sharif.edu}

 \author{Javad T. Firouzjaee}
\affiliation{School of Astronomy, Institute for Research in Fundamental Sciences (IPM), P. O. Box 19395-5531, Tehran, Iran }
 \email{j.taghizadeh.f@ipm.ir}
\author{Reza Mansouri}
\affiliation{Department of Physics, Sharif University of Technology,
Tehran, Iran and \\
  School of Astronomy, Institute for Research in Fundamental Sciences (IPM), P. O. Box 19395-5531, Tehran, Iran}
 \email{mansouri@ipm.ir}

\begin{abstract}
We study the time evolution of the Misner-Sharp mass and the apparent horizon for gravitational collapse of a massless scalar field in the $AdS_{5}$  space-time
for both cases of narrow and broad waves by numerically solving the Einstein's equations coupled to a massless scalar field. This is done by relying on the full dynamics of the collapse including the concept of the dynamical horizon. It turns out that the Misner-Sharp
mass is everywhere constant except for a rapid change across a thin shell defined by the density profile of the collapsing wave. By studying the evolution of the apparent horizon, indicating the formation of a black hole at different times we see how asymptotically an event horizon forms. The dependence of the thermalization time on the radius of the initial black hole event horizon is also studied.
\end{abstract}
%
%
\maketitle
\section{introduction}
There are two main motivations to study AdS and asymptotically
AdS black hole space times. First, the the AdS/CFT conjecture according to which there is a correspondence between the phenomena in AdS and those in a conformal field theory \cite{Maldacena}, \cite{Witten}; second, the AdS instability in the nonlinear regime \cite{Oscar J.C, P. Bizon, J. Jalmuzna, Yaffe} which ensures that a black hole is always formed regardless of the amplitude of perturbations. In this way, the AdS/CFT correspondence was extended to include field theories at finite temperature \cite{E. Witten}. In this correspondence a field theory in equilibrium at finite temperature is dual to an asymptotically AdS black hole. The non-equilibrium dynamics and the evolution towards thermalization in the field theory correspondes to the formation of a dynamical black hole horizon \cite{Balasubramanian, Bhattacharyya, Garfinkle, Bin Wu}; a concept which is not widely studied in general relativity \cite{ashtekar02}. \\
Now, we are interested on the gravity side both in the dynamics of the formation of a black hole within an AdS due to the instabilities, and in the dynamics of the state of an asymptotically AdS black hole after being perturbed; specifically we study the formation of the apparent horizon and its evolution towards the final state and the quasi-local mass one may assign to the dynamical black hole. Non of the problems have been tackled so far in the literature. To this aim, we need to model a dynamical black hole within an AdS background. Now, Ashtekar and Krishnan have presented a well-defined quantity for the dynamical black hole boundary and its area law which can be applied to a non-flat background such as FRW \cite{taghizadeh} or AdS. Attempts to build black holes in dynamical contexts have neglected these notions \cite{ads-bh}. Note that the black hole radiation temperature being crucial in the CFT thermalization temperature is attributed to the dynamical apparent horizon and not to the event horizon \cite{radiation, Das}. This has not been emphasized in \cite{Maeda}. Though this approach provides us with a well defined dynamical black hole concept, the notion of the mass however is neither trivial nor unique \cite {taghizadeh-mass}. 
  
Motivated by the fact that a four dimensional conformal field theory is dual to gravity on $AdS_{5}$, we study the dynamical case of the collapse of a massless scalar field coupled to gravity in the $AdS_{5}$. Such a collapse of a scalar field in the AdS background has been recently studied in \cite{Garfinkle, Bin Wu, Bhattacharyya, Maliborskis, J. Jalmuzna, scalar-collapse}. The results of our paper may then be used to study the corresponding  4-dimensional CFT features. Using the well developed approach to dynamical black hole concepts, we show that an apparent horizon will form and its formation time is affected by the presence of a black hole. The source amplitude will also affect the apparent horizon. In the dual description it means that injecting more energy into the boundary field theory will increase its temperature. In addition, we also show that thermalization time increases as the radius of the initial black hole increases. This is interpreted as if it will take longer for the boundary field theory with less temperature to reach another equilibrium state. Another dynamical notion is the concept of quasi-local mass. One such mass is Misner-Sharp mass \cite{Misner}. The Misner-Sharp mass gives us a clear picture of how different regions reach equilibrium: Regions near the boundary will reach equilibrium faster. Thermalized regions have a very flat Misner-Sharp mass behavior. We note that the Misner-Sharp mass for a static black hole coincides with the black hole mass.\\ 

In section II, we present the metric and the equations of motion in addition to the details of the numerics involved. Section III is devoted to different aspects of the dynamical
behavior of the model such as the mass and energy density, considering both narrow and broad waves. In the case of broad waves, a two stage collapse solution is found similar to \cite{Bin Wu}. In the case of narrow waves, two sets of initial conditions which are discussed are AdS vacuum and AdS with initial black hole. Quasi-local Misner-Sharp mass for this space time is presented and its evolution is studied by plotting its
behavior for different times. It is shown that the space is asymptotically AdS. It
is also demonstrated that the space time is divided into three regions. In section IV, we plot the evolution of metric functions to show how the black hole forms. By
evolving the equations for longer times we study apparent horizon formation. We show that it approaches the AdS black hole horizon as one
expects. Different quantities of interest such as event horizon radius and black hole temperature are computed. We then conclude in section V.

\section{Dynamical metric for the  AdS background}
Thermalization of a spatially homogeneous system which starts initially out of equilibrium by a scalar source coupled to a marginal operator corresponds on the gravity side to the collapse of a massless scalar field in the AdS space time. Now, the value of the (boundary) source $\varphi_{0}$, is given by the value of the five dimensional field at the boundary of AdS space time. We, therefore, start with the Einstein-Hilbert action in $AdS_{5}$ with a minimally coupled massless scalar field written as
\begin{equation}
S=\frac{1}{2k_{4+1}^2} {\int d^{4+1}x \sqrt{-g}(R-2\Lambda -2(\partial \varphi)^{2})}.
\label{1}
\end{equation}

For a spatially homogeneous system on the boundary, the metric form in the Poincaré coordinates is given by
\begin{equation}
ds^2=\frac{1}{u^2}(-fe^{\delta} dt^2+\frac{1}{f}du^2+dx^2),
\label{2}
\end{equation}
where $f$ and $\delta$ are functions of $t$ and $u$.
The AdS vacuum solution corresponds to $\delta=0$ and $f=1$, while for the AdS black hole we set $f=1-\frac{u^4}{u_{0}^4}$, $\delta=0$ and $\frac{1}{u_{0}}=\pi T$ with $T$ being the black hole temperature.

In this coordinate system, Einstein-Klein-Gordon equations reduce to the following set of differential equations:
\begin{equation}
\dot{V}=u^{3}(\frac{fe^{-\delta} P}{u^3})^{'} ,\\
\label{3}
\end{equation}
\begin{equation}
\dot{P}=(fe^{-\delta}V)^{'} ,\\
\label{4}
\end{equation}
\begin{equation}
\dot{f} =\frac{4}{3}uf^2 e^{-\delta}VP ,\\
\label{5}
\end{equation}
\begin{equation}
\delta ^{'}=\frac{2}{3}u(V^2+P^2),\\
\label{6}
\end{equation}
and
\begin{equation}
f^{'}=\frac{2}{3} f(V^2+P^2)+\frac{4}{u}(f-1),
\label{7}
\end{equation}
where $ P=\varphi'$,
$ V=f^{-1} e^{\delta} \dot{\varphi}$, and the derivatives with respect to t and u are denoted by overdots and primes. The last equation is a constraint equation. We have therefore a set of differential equations defining an initial value problem. The initial conditions and boundary conditions are presented in the following subsection.

\subsection{Initial and boundary conditions and the solution algorithm}
To solve the above equations one needs to impose initial and boundary conditions. We are interested in both narrow waves and broad waves. In the case of narrow waves two different sets of initial conditions for the vacuum AdS and the AdS black hole are considered. For broad waves, however, only the vacuum AdS initial condition is considered.
In the case of the vacuum AdS initial condition we have
\begin{equation}
f=1, P=V=\delta = 0,
\end{equation}
and for the case of the AdS black hole initial condition we set
\begin{equation}
f=1-u^{4},P=V=\delta =0.
\end{equation}
The boundary condition can not be given at $u = 0$ due to the choice of the coordinates making the system of equations singular at $u=0$. Therefore, we set a lower bound for the numerical calculation, i.e $u_{min}=0.005$.
Now, to solve the equation (\ref{6}) we fix the boundary condition by
\begin{equation}
\delta(t, u_{min})=0 .
\end{equation}
The $\varphi$ value at the boundary, in the dual picture, injects energy in to the boundary field \cite{de Haro:2000xn}.
To study thermalization we require that systems of equations reach an equilibrium. This can be achieved if the source vanishes effectively after some time interval.
 The boundary conditions for the equation (\ref{3}), (\ref{4}) are
\begin{equation}
V(t,0)=-2t\epsilon e^{-at^2}, P(t,\infty)=0.
\label{A1}
\end{equation}
In order to solve the equations numerically, a cut off  $u_{max}$ is needed. Given the fact that there is always a region in which the wave vanishes, we may select
any large value for $u_{max}$ within that region (Fig.\ref{fig:1}). We then have used the third order Adams-Bashforth method for time-marching and finite difference
in the spatial part \cite{Bin Wu}, denoting by $dt$ the time intervals and by $du$ the space intervals. The value of each function at the
point $(t_{i},u_{j})$ is then written as $h^{i}_{j}$ in which $t_{i}=t_{0}+idt$ and $u_{j}=u_{min}+jdu$. The algorithm is checked to be robust against the change
of the step size as far as the time steps are less than half of the space steps, the steps are not too big, and the space time approaches AdS on the boundary. Now,
for each function using the third order Adams-Bashforth method we have
\begin{equation}
h_{j}^{i+1}=h_{j}^{i}+\frac{dt}{12}(23 {\dot{h}}^{i}_{j}-16\dot{h}^{i-1}_{j}+5\dot{h}^{i-2}_{j} ).
\label{AB}
\end{equation}
Thus, the system of differential equations are given by
\begin{eqnarray}
\dot{V}_{j}^{i}=-\frac{3}{u_{j}}(f^{i}_{j} e^{-\delta^{i}_{j}} P^{i}_{j})+\frac{f_{j}^{i}e^{\delta^{i}_{j}}P^{i}_{j}-f_{j-1}^{i}e^{\delta^{i}_{j-1}}P^{i}_{j-1}}{du}\\
\dot{P}^{i}_{j}=\frac{(f^{i}_{j+1}e^{-\delta_{j+1}^{i}}V^{i}_{j+1}-f^{i}_{j}e^{-\delta_{j}^{i}}V^{i}_{j})}{du}\\
\dot{f}^{i}_{j}=\frac{4}{3}u_{j}{f^{i}_{j}}^{2}e^{-\delta^{i}_{j}}V^{i}_{j}P^{i}_{j}\\
\frac{\delta^{i}_{j}-\delta^{i}_{j-1}}{du}=\frac{2}{3}u_{j}({V^{i}_{j}}^{2}+{P^{i}_{j}}^2).
\end{eqnarray}

 Given the initial conditions, at each time interval $t_{n}$, we first compute $f,V,P$  using equation (\ref{AB}), before going to $\delta $. With this conditions, the value
  $ \dot P, \dot V$ and $ \dot f$
  are computed using the finite difference method for each hyper-surface.


\section{Energy and mass behavior}

\begin{figure}[htbp!]
\centering
\includegraphics[width=1 \columnwidth]{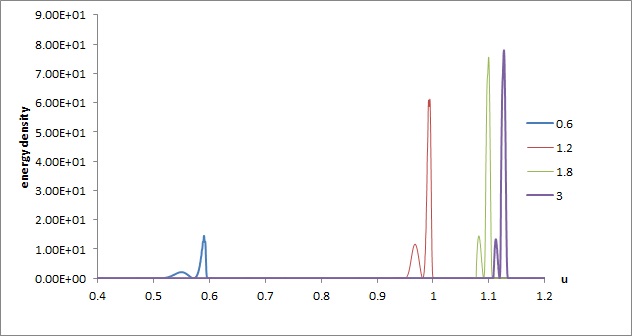}
\caption{Density evolution of the collapsing wave with the vacuum AdS initial condition. The wave becomes more compact as it propagates inward.}
\label{fig:2}
\end{figure}

\begin{figure}[htbp!]
\centering
\includegraphics[width=1 \columnwidth]{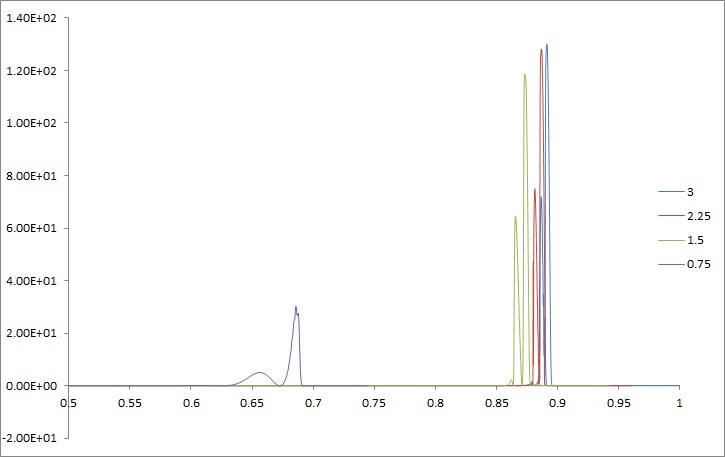}
\caption{Density evolution of the collapsing wave with the AdS black hole initial condition.}
\label{fig:2-2}
\end{figure}

We study both cases with the source turned on for short $(\frac{\varepsilon}{a}\leq 1, \Delta t<\frac{1}{T})$ and long $(\frac{\varepsilon}{a}\geq 1, \Delta t > \frac{1}{T})$ durations. The limit in which $\dfrac{\varepsilon}{a}\ll1$ induces a narrow wave near the boundary.
As stated, in the case of narrow waves two sets of initial conditions are considered, namely for the vacuum AdS and the AdS black hole. The results for both cases are presented.
We choose $(a,\varepsilon)=(400,0.5)$ for the parameters. This produces a wave with the duration $\Delta t=0.05$. The source is then turned off and an ingoing narrow wave is produced.
The energy density for this space time is defined as $f(V^{2}+P^{2})$. Fig.\ref{fig:2} and Fig.\ref{fig:2-2} show energy density of the field for AdS and AdS black hole initial conditions, respectively. As a test of our numerics, we note similarity of our results for the AdS case with those reported in \cite{Bin Wu}. From the Figs one understands that as the wave propagates inward it becomes more compact with the density becoming sharply peaked before it collapses and the apparent horizon forms. Formation of the apparent horizon is best seen by studying the behavior of the metric function $f$ which is presented in section IV.
Concepts like black hole and its horizon, dynamical horizon, and the mass of a compact over-density region are the main features of a dynamical system within gravity which help us to understand concepts like equilibrium within the context of AdS/CFT. 
The concept of mass, however, is obscured by the fact that in dynamical settings in general relativity we do not have a well defined concept of mass and energy. Instead we can attribute mass to compact surfaces leading to many different definitions coined as the quasi-local mass. The Misner-Sharp quasi-local mass is the one having significant importance in thermodynamics of black holes reducing to the Newtonian concept of mass \cite{taghizadeh}.\\
 Assume any $n$ dimensional manifold of $\mathcal{M}\approx m^{2}*K^{n-2}$ with the line element given by
\begin{equation}
ds^{2}=g_{AB}dy^{A}dy^{B}+{R(y)}^{2}\gamma_{ij}(z)dz^{i}dz^{j},
\end{equation}
where $K^{n-2}$ is a $n-2$ dimensional space of constant curvature. The quasi-local Misner-Sharp mass is defined as \cite{Maeda}
\begin{equation}
M(t,\overrightarrow{x})=\frac{(n-2)V^{k}_{n-2}}{2k_{n}^{2}}R^{n-3}\lbrace-\widehat{\Lambda}R^{2}+k-(DR)^{2}\rbrace,
\end{equation}
where $g_{AB}$ is a Lorentzian metric on $ m^{2}$, $D_{A}$ is the covariant derivative on $ m^{2}$, $V^{k}_{n-2}$ is the area of the $n-2$ dimensional part and $\widehat \Lambda=\frac{2\Lambda}{(n-1)(n-2)}$, where $\Lambda$ is the cosmological constant. In our case we have set $ \widehat \Lambda=1$, $n=5$ and $k=0$.  
The quasi-local Misner-Sharp mass for this space time is computed as
\begin{equation}
 M(t,u)=\frac{1-f}{u^4} \cdot
\end{equation}

\begin{figure}[htbp!]
\centering
\includegraphics[width=1 \columnwidth]{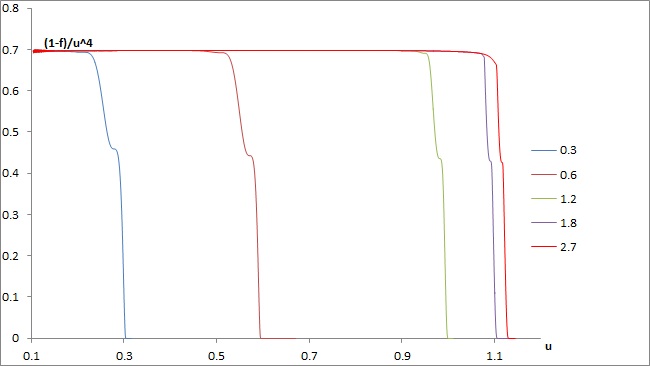}
\caption{Time evolution of Misner-Sharp mass with vacuum AdS initial condition. The mass decreases from a constant value outside the wave to zero after going through a transition region.}
\label{fig:4}
\end{figure}

\begin{figure}[htbp!]
\centering
\includegraphics[width=1 \columnwidth]{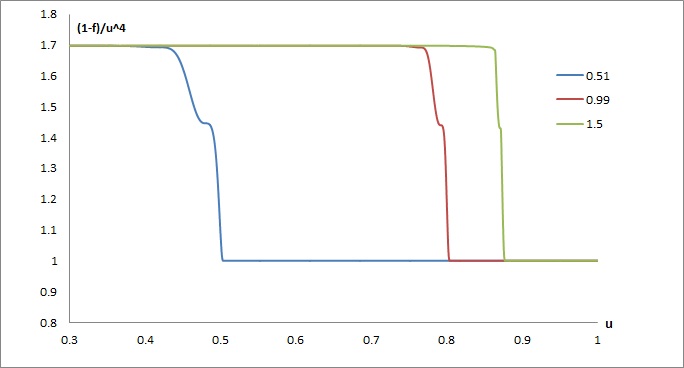}
\caption{Time evolution of Misner-Sharp mass with AdS black hole initial condition. Note the behavior of the mass in three regions similar to the vacuum AdS initial condition }
\label{fig:4-4}
\end{figure}

We use the Misner-Sharp mass, being a constant for the AdS-Schwarzschild solution, as a measure of how different regions reach equilibrium. From Fig.\ref{fig:4} and Fig.\ref{fig:4-4} we may differentiate three different parts of the space time: $(1)$ a static AdS-Schwarzschild region behind the wave corresponding to a thermal equilibrium in the corresponding CFT; $(2)$ a narrow transition region equivalent to a non-equilibrium state; $(3)$ the vacuum AdS. As elaborated in \cite{Balasubramanian}, we may say that in the corresponding CFT description thermalization happens from small to large scales.

\begin{figure}[htbp!]
\centering
\includegraphics[width=1 \columnwidth]{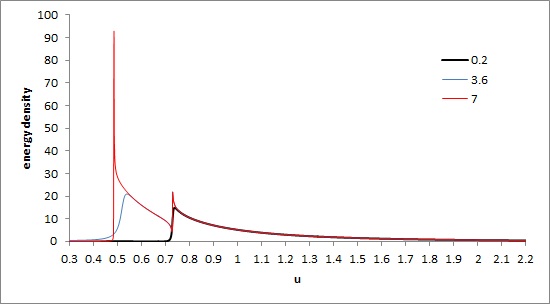}
\caption{Density evolution in the case of broad waves. Two peaks will start to evolve leading to a two stage collapse. }
\label{fig:4-4-4}
\end{figure}

The limit $\dfrac{\varepsilon}{a}\geqslant 1$ induces a broad wave at the boundary. If the duration of the source is long enough $(\Delta t>\frac{1}{T})$ this will lead to a two stage collapse. We choose $(a,\varepsilon)=(0.1,0.3)$ for the parameters of the wave. By starting from negative times, the energy is transferred to the bulk by two pulses of the source according to the equation (\ref{A1}). This energy density at different times is plotted in Fig.\ref{fig:4-4-4}. Two peaks are observed in the energy density corresponding to two collapses.

\section{Black hole formation }

\begin{figure}[htbp!]
\centering
\includegraphics[width=1 \columnwidth]{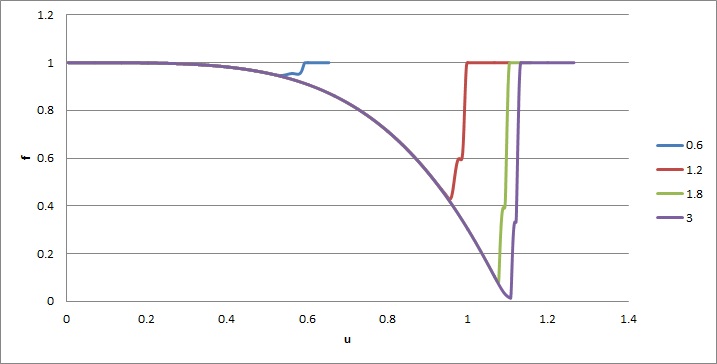}
\caption{The evolution of the metric factor $f$ in the case of vacuum AdS initial condition. As the wave becomes compact it starts to collapse. This will result in the formation the apparent horizon.}
\label{fig:1}
\end{figure}

\begin{figure}[htbp!]
\centering
\includegraphics[width=1 \columnwidth]{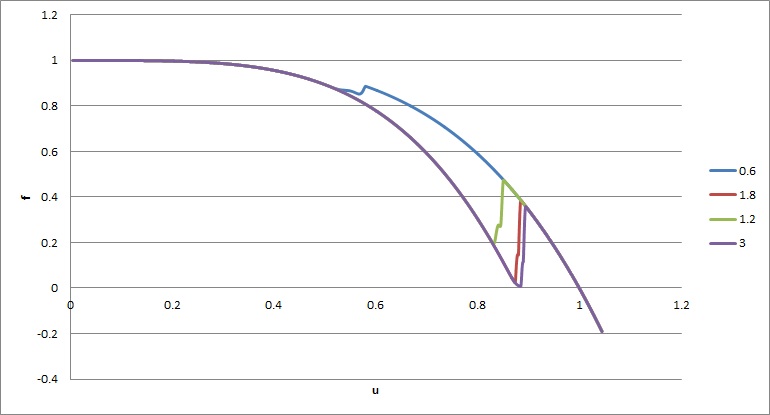}
\caption{The evolution of the metric factor $f$ in the case of AdS black hole initial condition. The apparent horizon formation time decreases compared to the vacuum AdS initial condition.}
\label{fig:1-1}
\end{figure}

Let's now go over to the question of if and how a black hole can form during the collapse of the ingoing wave. Within this dynamical setting the concept of a static black hole defined by the existence of an event horizon can not be used. We therefore turn to the concept of the apparent horizon applicable in any dynamical setting. To define the apparent horizon being a $d-1$ dimensional hypersurface in a $d$ dimensional space-time we need first the induced metric given by
\begin{equation}
\tilde{q_{ab}}= g_{ab}+\ell_{a} n_{b}+n_{a} \ell_{b},
\end{equation}
where $n^a$ and $\ell^b$ are ingoing and outgoing geodesics normal to the $d-1$ dimensional hyper-surface. Two scalar quantities called geodesic expansions are defined by
\begin{eqnarray}
\theta_\ell=\tilde{q^{ab}} \nabla_a \ell_b ,
\end{eqnarray}

\begin{eqnarray}
\theta_n=\tilde{q^{ab}} \nabla_a n_b.
\end{eqnarray}
The region in which the expansions $\theta_\ell$ and $\theta_n$ are negative is called the trapped region and its boundary is the apparent horizon. Now, in our case
the $f=0$ hypersurface is the boundary of the trapped region, or the apparent horizon, where the expansion of the outgoing geodesics is zero. However, to calculate
numerically, we set a lower bound for $f$, say $f = 0.02$. The behavior of the metric function $f$ for the narrow wave in the case of AdS for all regions of space-time is shown in Fig.\ref{fig:1}. Note that the value of $f$ is almost equal to to one outside and inside the wave, and has a minimum which decreases to zero ($0.02$) once the black hole's apparent horizon is formed. The case of AdS black hole is shown in Fig.\ref{fig:1-1}. We notice a small increase of $f$ after going down to zero due to the existence of a gap between collapsing wave and the central black hole. Starting with vacuum AdS initial condition the collapse occurs at $u\sim 1.1$. If we define the apparent horizon formation time $t_{A}$ such that $f$ is approximately less than $0.02$ then we find $t_{A}\sim 2.3 $. Given the initial condition of the AdS black hole, the collapse occurs at $u\sim 0.9$ and $t_{A}\sim1.7$. Therefore, given the same parameters of the boundary source, in the presence of an initial black hole the formation time of the apparent horizon decreases. 

\begin{figure}[htbp!]
\centering
\includegraphics[width=1 \columnwidth]{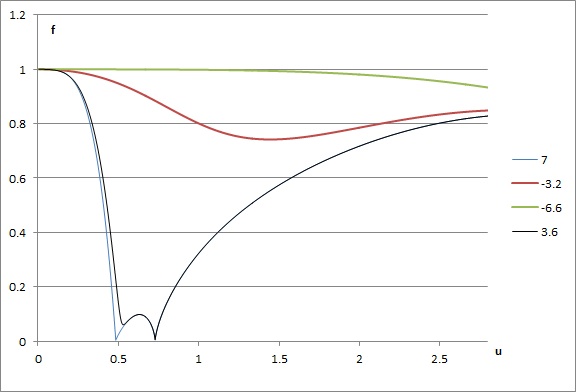}
\caption{The evolution of the metric factor $f$ in the case of broad waves. Two pulses will start to collapse leading to a two stage collapse. }
\label{fig:1-1-1}
\end{figure}

The case of broad waves is shown in Fig.\ref{fig:1-1-1}, where we see the formation of a two phase black hole. The first collapse happens at $u\sim 0.7$ and
the second one at $u\sim 0.5$.

\subsection{The horizon behavior }
\begin{figure}[htbp!]
\centering
\includegraphics[width=1 \columnwidth]{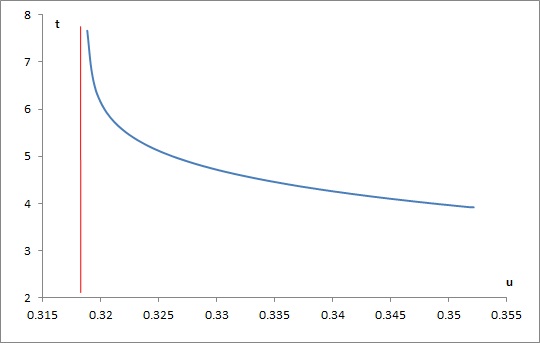}
\caption{The apparent horizon reaches the event horizon at late time.}
\label{fig:5a}
\end{figure}

\begin{figure}[htbp!]
\centering
\includegraphics[width=1 \columnwidth]{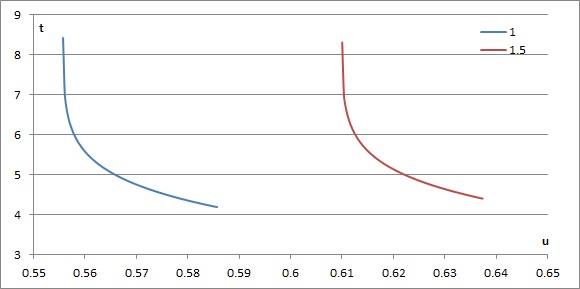}
\caption{The formation of the apparent horizon for different initial black holes.}
\label{fig:5b}
\end{figure}

\begin{figure}[htbp!]
\centering
\includegraphics[width=1 \columnwidth]{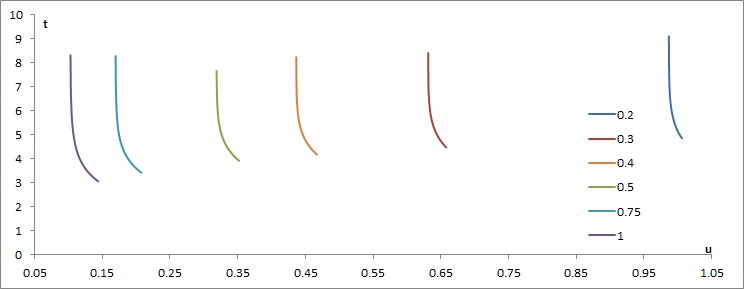}
\caption{The place of the event horizon depends on the amplitude of the collapsing wave.}
\label{fig:5}
\end{figure}

The collapse of a wave in an AdS-Schwarzschild black hole ends asymptotically into an AdS-Schwarzschild black hole again, irrespective of the wave amplitude. This is seen by the following argument. During the collapse of the incoming wave the apparent horizon is defined by the boundary of the trapped region leading to $f(t,u) = 0$ for our models. This is done numerically by setting the amplitude of the
boundary source to $\epsilon=0.5$. Fig.\ref{fig:5a} shows the behavior of the apparent horizon, approaching asymptotically the event horizon. This is a typical behavior independent of the initial black hole radius and wave amplitude $\varepsilon$, as may be seen from Fig.\ref{fig:5b} and Fig.\ref{fig:5}.
Note that the radius of the resulting black hole and the time of its formation decreases as the amplitude of the wave increases.
The temperature of the black hole is given by
$T=\frac{1}{\pi u_{0}}$, where the $u_{0}$ defines the apparent horizon radius at late time \cite{Das}.
Therefore, the temperature of the black hole increases as the amplitude increases.
 The robustness of the algorithm is also checked in this case. We may note that in the case of too big time and space steps the boundary conditions may not be maintained.

\subsection{Black hole thermalization }

\begin{figure}[htbp!]
\centering
\includegraphics[width=1 \columnwidth]{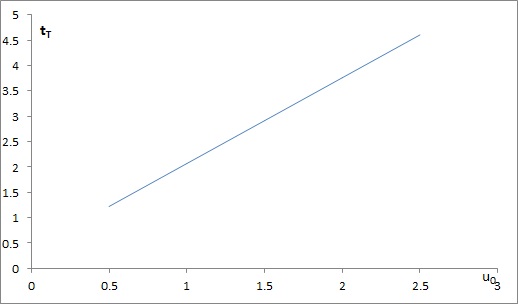}
\caption{The thermalization time depends on the initial black hole radius.}
\label{fig:y}
\end{figure}

We define the thermalization time as the time in which $f$ reaches a minimum value of $0.01$. In Fig.\ref{fig:y} we have depicted the
dependence of the thermalization time $t_{T}$ on the radius of the initial black hole event horizon, for the same parameters of the
wave packet. 
In this coordinate the mass of the black hole is given by $M=\frac{1}{{u_{0}}^{4}}$. Therefore, the thermalization time increases with the decrease
of the black hole mass. It will also take longer for states with less temperature to reach another equilibrium state.

\section{Conclusion }

Studying the dynamical aspects of the AdS and AdS black hole instabilities has been the goal of the research communicated in this paper. Not only the knowledge of the final state but the dynamics of the collapse of a perturbation wave within such space times and the formation of the apparent horizon leading to an event horizon, and the time evolution of the corresponding  Misner-Sharp mass has been the subject of our study. For the sake of having useful results in a 4-dimensional conformal field theory in the AdS/CFT context, we have looked at the dynamics of the collapse of a massless scalar field in the five dimensional AdS and Schwarzschild-AdS space-time, and developed a robust numerics to calculate different aspects of the dynamics, such as the features of the apparent horizon and quasi-local Misner-Sharp mass.\\ 
We first showed that an apparent horizon will form which asymptotically becomes an event horizon. We also showed that the formation time of the apparent horizon decreases
as the source amplitude increases. This means that in the corresponding field theory injecting more energy will result in a higher temperature.
It was also noted that the apparent horizon formation time is delayed in vacuum AdS case compared to AdS with an initial black hole.
 Furthermore, the effect of initial AdS black hole radius on the apparent horizon formation time is considered. We concluded that a decrease in the initial AdS black hole radius will decrease the formation time of the apparent horizon.
It has also been shown that the presence of an initial black hole affects the thermalization time. Using the thermalization definition given in \cite{Bin Wu}, we found that the increase of the radius of the initial black hole increases the thermalization time.\\

In addition, by studying Misner-Sharp mass we concluded how different regions of space-time reach an equilibrium state. Regions near the boundary of the wave reached equilibrium faster. In the boundary field theory this means that small scales thermalize faster. We showed that Misner-Sharp mass at different times turns out to be almost constant
except for the narrow transition region identifying three different regions of the space-time. 

\section{acknowledgment}

We would like to thank Mojahed Parsi Mood for helpful discussion .


\begin{thebibliography}{99}

\bibitem{Maldacena}
J.~M.~Maldacena,
  Adv.\ Theor.\ Math.\ Phys.\  {\bf 2}, 231 (1998)
  [hep-th/9711200].

\bibitem{Witten}
 E.~Witten,
  Adv.\ Theor.\ Math.\ Phys.\  {\bf 2}, 253 (1998)
  [hep-th/9802150].


\bibitem{Garfinkle}
D.~Garfinkle, L.~A.~Pando Zayas and D.~Reichmann,
  JHEP {\bf 1202}, 119 (2012)
  [arXiv:1110.5823 [hep-th]].

\bibitem{P. Bizon}
 P.~Bizon and A.~Rostworowski,
  Phys.\ Rev.\ Lett.\  {\bf 107}, 031102 (2011)
  [arXiv:1104.3702 [gr-qc]].

\bibitem{Alex Buche}
A.~Buchel, S.~L.~Liebling and L.~Lehner,
  Phys.\ Rev.\ D {\bf 87}, 123006 (2013)
  [arXiv:1304.4166 [gr-qc]].

\bibitem{Bin Wu}
B.~Wu,
  JHEP {\bf 1210}, 133 (2012)
  [arXiv:1208.1393 [hep-th]].

\bibitem{Oscar J.C}
O.~J.~C.~Dias, G.~T.~Horowitz and J.~E.~Santos,
  Class.\ Quant.\ Grav.\  {\bf 29}, 194002 (2012)
  [arXiv:1109.1825 [hep-th]].

\bibitem{Bhattacharyya}
S.~Bhattacharyya and S.~Minwalla,
  JHEP {\bf 0909}, 034 (2009)
  [arXiv:0904.0464 [hep-th]].


\bibitem{Balasubramanian}
V.~Balasubramanian, A.~Bernamonti, J.~de Boer, N.~Copland, B.~Craps, E.~Keski-Vakkuri, B.~Muller and A.~Schafer {\it et al.},
  Phys.\ Rev.\ D {\bf 84}, 026010 (2011)
  [arXiv:1103.2683 [hep-th]].

\bibitem{E. Witten}
E.~Witten,
  Adv.\ Theor.\ Math.\ Phys.\  {\bf 2}, 505 (1998)
  [hep-th/9803131].

\bibitem{Maliborskis}
M.~Maliborski and A.~Rostworowski,
  arXiv:1307.2875 [gr-qc].

\bibitem{M. Maliborski}
 M.~Maliborski and A.~Rostworowski,
  Phys.\ Rev.\ Lett.\  {\bf 111}, 051102 (2013)
  [arXiv:1303.3186 [gr-qc]].


\bibitem{J. Jalmuzna}
J.~Jalmuzna, A.~Rostworowski and P.~Bizon,
  Phys.\ Rev.\ D {\bf 84}, 085021 (2011)
  [arXiv:1108.4539 [gr-qc]].

\bibitem{De Oliveira}
  H.~P.~de Oliveira, L.~A.~Pando Zayas and E.~L.~Rodrigues,
  Phys.\ Rev.\ Lett.\  {\bf 111}, 051101 (2013)
  [arXiv:1209.2369 [hep-th]].
\bibitem{ashtekar02}
  A.~Ashtekar and B.~Krishnan,
  Phys.\ Rev.\ Lett.\  {\bf 89}, 261101 (2002)
  [gr-qc/0207080].

\bibitem{taghizadeh}
J.~T.~Firouzjaee and Reza~Mansouri,
  Gen.\ Rel.\ Grav.\  {\bf 42}, 2431 (2010)
  [arXiv:0812.5108 [astro-ph]];
  Rahim~Moradi, J.~T.~Firouzjaee and Reza~Mansouri,
  arXiv:1301.1480 [gr-qc].

\bibitem{Maeda}
H.~Maeda,
  Phys.\ Rev.\ D {\bf 86}, 044016 (2012)
  [arXiv:1204.4472 [gr-qc]].
\bibitem{ads-bh}
 Nakao K,
Gen Rel. Grav. 24 (1992) 1069-1081;
J.~P.~S.~Lemos,
  Phys.\ Rev.\ D {\bf 59}, 044020 (1999);  Markovic D and Shapiro S L,
Phys. Rev.61 (2000) 084029; Lake K,
 Phys. Rev. D 62 (2000), 027301; S.~B.~Giddings and A.~Nudelman,
  JHEP {\bf 0202}, 003 (2002)
  [hep-th/0112099].

\bibitem{taghizadeh-mass}
J.~T.~Firouzjaee, M.~Parsi~Mood and Reza~Mansouri,
  Gen.\ Rel.\ Grav.\  {\bf 44}, 639 (2012)
  [arXiv:1010.3971 [gr-qc]].

\bibitem{radiation}
M.~Visser,
  Int.\ J.\ Mod.\ Phys.\ D {\bf 12}, 649 (2003)
  [hep-th/0106111]
; J.~T.~Firouzjaee and Reza~Mansouri,
  Europhys.\ Lett.\  {\bf 97}, 29002 (2012)
  [arXiv:1104.0530 [gr-qc]].

 \bibitem{scalar-collapse}
A.~Buchel, L.~Lehner and S.~L.~Liebling,
  Phys.\ Rev.\ D {\bf 86}, 123011 (2012)
  [arXiv:1210.0890 [gr-qc]]
; B.~Wu,
  JHEP {\bf 1304}, 044 (2013)
  [arXiv:1301.3796 [hep-th]]
; D.~Garfinkle and L.~A.~Pando Zayas,
  Phys.\ Rev.\ D {\bf 84}, 066006 (2011)
  [arXiv:1106.2339 [hep-th]].

\bibitem{de Haro:2000xn}
  S.~de Haro, S.~N.~Solodukhin and K.~Skenderis,
  Commun.\ Math.\ Phys.\  {\bf 217} (2001) 595
  [hep-th/0002230].


\bibitem{Yaffe}
  P.~M.~Chesler and L.~G.~Yaffe,
  arXiv:1309.1439 [hep-th].

\bibitem{Das}
  S.~R.~Das, T.~Nishioka and T.~Takayanagi,
  JHEP {\bf 1007}, 071 (2010)
  [arXiv:1005.3348 [hep-th]].
\bibitem{Misner}
C. W. Misner and D. H. Sharp, Phys. Rev. 136 (1964)
B571.

\end{thebibliography}
\end{document}